\documentclass[prl,twocolumn,superscriptaddress,nopacs,nofootinbib]{revtex4}
\usepackage{graphicx}

\usepackage{amsmath}
\usepackage{amssymb}
\usepackage{accents}
\usepackage{color}
\usepackage{verbatim}
\usepackage{hyperref}
\usepackage{empheq} 
\definecolor{very-light-gray}{gray}{0.9}

\begin{document}

\title{Slow non-thermalizing dynamics in a quantum spin glass}
\author{Louk Rademaker}
\author{Dmitry A. Abanin}
\affiliation{Department of Theoretical Physics, University of Geneva, 1211 Geneva, Switzerland}

\date{\today}

\begin{abstract} 
Spin glasses and many-body localization  (MBL) are prime examples of ergodicity breaking, yet their physical origin is quite different: the former phase arises due to rugged classical energy landscape, while the latter is a quantum-interference effect. Here we study quantum dynamics of an isolated 1d spin-glass under application of a transverse field.  At high energy densities, the system is ergodic, relaxing via resonance avalanche mechanism, that is also responsible for the destruction of MBL in non-glassy systems with power-law interactions. At low energy densities, the interaction-induced fields obtain a power-law soft gap, making the resonance avalanche mechanism inefficient. This leads to the persistence of the spin-glass order, as demonstrated by resonance analysis and by numerical studies. A small fraction of resonant spins forms a thermalizing system with long-range entanglement, making this regime distinct from the conventional MBL. The model considered can be realized in systems of trapped ions, opening the door to investigating slow quantum dynamics induced by glassiness.
\end{abstract}

\maketitle

{\bf Introduction.} Spin glasses (SG) and many-body localization (MBL) are two broad classes of systems that break ergodicity. A SG is a system where frustration resulting from random interactions and fields cause spins to `freeze' at low temperatures, leading to an `ordered' phase without long-range order~\cite{mezard1987spin,mydosh1993spin,stein2013spin}. Characteristic extremely slow dynamics in SG originates from a large number of metastable low-energy states separated by large energy barriers. 

While SG models are essentially classical, the emergence of freezing in quantum systems is being actively studied theoretically and experimentally in the context of MBL~\cite{Nandkishore:2015kt,AbaninRMP}. Although, as with SG, randomness is essential, the fundamental mechanism behind MBL is quantum interference, rather than frustration.

Another major discrepancy between SG and MBL immediately meets the eye: MBL exists in short-range interacting models in $d=1$-dimensional systems~\cite{Nandkishore:2015kt,AbaninRMP}; whereas SG require long-range interactions or more than $d=2$ dimensions. Moreover, analytical results for SG exist predominantly for models with {\it infinite}-range interactions~\cite{mezard1987spin}. There are of course some exceptions, notably, a $d=1$-dimensional long-range interacting spin glass model introduced by Kotliar, Anderson and Stein~[\onlinecite{Kotliar:1983fo}]. 

The question regarding similarities and differences in dynamics between spin glasses and many-body localized phases remains largely open. Recently, eigenstates~\cite{PalSpinGlass,Burin:2017ds,Mukherjee:2018jz} and dynamics~\cite{LeticiaSG,SchiroSpinGlass2019} of {\it infinite-range} spin glasses have been studied. Such models are difficult to realize (see, however, Ref.~\cite{PhysRevLett.107.277202}); here, instead our focus will be on experimentally relevant systems with power-law decaying interactions.

We propose to bridge the gap between spin glasses and MBL by studying the Kotliar-Anderson-Stein SG model~ [\onlinecite{Kotliar:1983fo}] with a quantum transverse field. This model has the advantage of being experimentally realizable. In particular, 1d disordered systems with long-range interactions have been recently studied with trapped ions~\cite{Smith:2016cd}. We investigate quench dynamics of this model at high and low (but nonzero) energy densities $\epsilon$, finding that the onset of glassiness dramatically modifies dynamics at low $\epsilon$. Throughout, we will focus on the properties of {\it isolated} systems; note that the dynamics of glasses in the presence of external bath has been investigated extensively~\cite{PARISI_LesHouches}.   

Recent works~\cite{Maksymov:2017dp,Burin:2015kz,Burin:2015gq, Gutman16,Yao14Dipolar} argued that MBL is impossible in the thermodynamic limit in 1d for sufficiently long-ranged power-law interactions, attributing numerical signatures of MBL reported in Refs.~\cite{Hauke:2015fm,Li:2016bg,Nag:2019gz} to finite-size effects. We argue that the novel aspect -- frustration and glassiness -- of our model compared to those studied in Ref.~\cite{Maksymov:2017dp,Burin:2015kz,Burin:2015gq, Gutman16,Yao14Dipolar} enables ergodicity breaking in the quantum model at low energy density. It is worth noting that Ref.~\cite{Nandkishore:2017dl} proposed that MBL may occur at low energy density in systems with long-range interactions via a very different mechanism of charge confinement.

{\bf Model and setup.} The Hamiltonian of the long-range quantum spin glass model of interest is given by:
\begin{equation}
	H = \sum_{ij} \frac{J_{ij}}{|i-j|^\alpha} \hat Z_i \hat Z_j - h_x \sum_i  \hat X_i
	\label{MainHamiltonian}
\end{equation}
where $\hat X_i, \hat Z_i$ are the Pauli operators for the spin on site $i$. Following Ref.~[\onlinecite{Kotliar:1983fo}], $J_{ij}$ are chosen to be random, normal distributed with standard deviation 1. All energies and times are therefore dimensionless. Parameter $\alpha$ sets the power of long-ranged interactions, and lies in the range $\frac{1}{2} < \alpha < 1$\footnote{Note that $\alpha=0$ does not correspond to the Sherrington-Kirkpatrick (SK) model, because in Eq.~\eqref{MainHamiltonian} the random variables $J_{ij}$ have a standard deviation independent of system size.} so that in the absence of a transverse field $h_x$ the system is SG at low temperatures~[\onlinecite{Kotliar:1983fo}]. Monte Carlo simulations of the classical model with $h_x = 0$ demonstrated a SG phase with critical temperature $T_c = 0.6$ for $\alpha = 0.75$~\cite{Katzgraber:2003cz}.

To probe the dynamical properties of the model~(\ref{MainHamiltonian}), we will focus on a quantum quench protocol, in which the system is initially prepared in a product state, with spins pointing along $z$-direction, $Z_i=\pm 1$. The quantity of interest is the decay of the initial magnetization pattern under unitary evolution with the Hamiltonian (\ref{MainHamiltonian}). This setup has been successfully used in cold atoms~\cite{Bloch15,Bloch16-2,GreinerMBLEntanglement} and trapped-ion~\cite{Smith:2016cd} experiments to probe ergodicity breaking via MBL. 

{\bf Effective fields.} In an initial product state, each spin is subject to a random $z$-field $\phi_i$ arising from the interaction term in Eq.~(\ref{MainHamiltonian}),
\begin{equation}
	\sum_{ij} \frac{J_{ij}}{|i-j|^\alpha} Z_i Z_j
	= \sum_i Z_i \phi_i, \; \; \; 
	\phi_i \equiv \sum_j \frac{J_{ij}}{|i-j|^\alpha} Z_j.
\end{equation}
For $h_x\ll 1$, a typical spin will have $h_x\ll |\phi_i|_{\rm typ}$, and such spin will at first precess around $z$-axis, maintaining the memory of its initial state. To relax and ``forget" its magnetization, the spin should either be involved in a higher-order multi-spin resonant process, or its on-site field $\phi_i$ should become smaller than $h_x$. Both processes are very sensitive to the distribution of on-site fields. 
We will first study this distribution, finding a markedly different behavior at high and low $\epsilon$, consistent with Ref.~\cite{Boettcher2008}.

At infinite temperature, meaning random, uncorrelated $Z_i=\pm 1$, the effective fields are normal distributed with standard deviation $\sigma_{T=\infty} = \sqrt{2 \zeta(2\alpha)}$, which diverges as $\alpha$ approaches 0.5 as $1/\sqrt{\alpha -1/2}$. Fields experienced by different spins are uncorrelated. 

In contrast, if we restrict ourselves to low energy density, the number of small fields gets suppressed. To prepare low energy configurations, we start with a random state and flip each spin aligned with the local field, until each spin is anti-aligned with the on-site field $\phi_i$. We call such states single-spin-flip stable (SFS).\cite{SupplInfo}

\begin{figure}
	\includegraphics[width=\columnwidth]{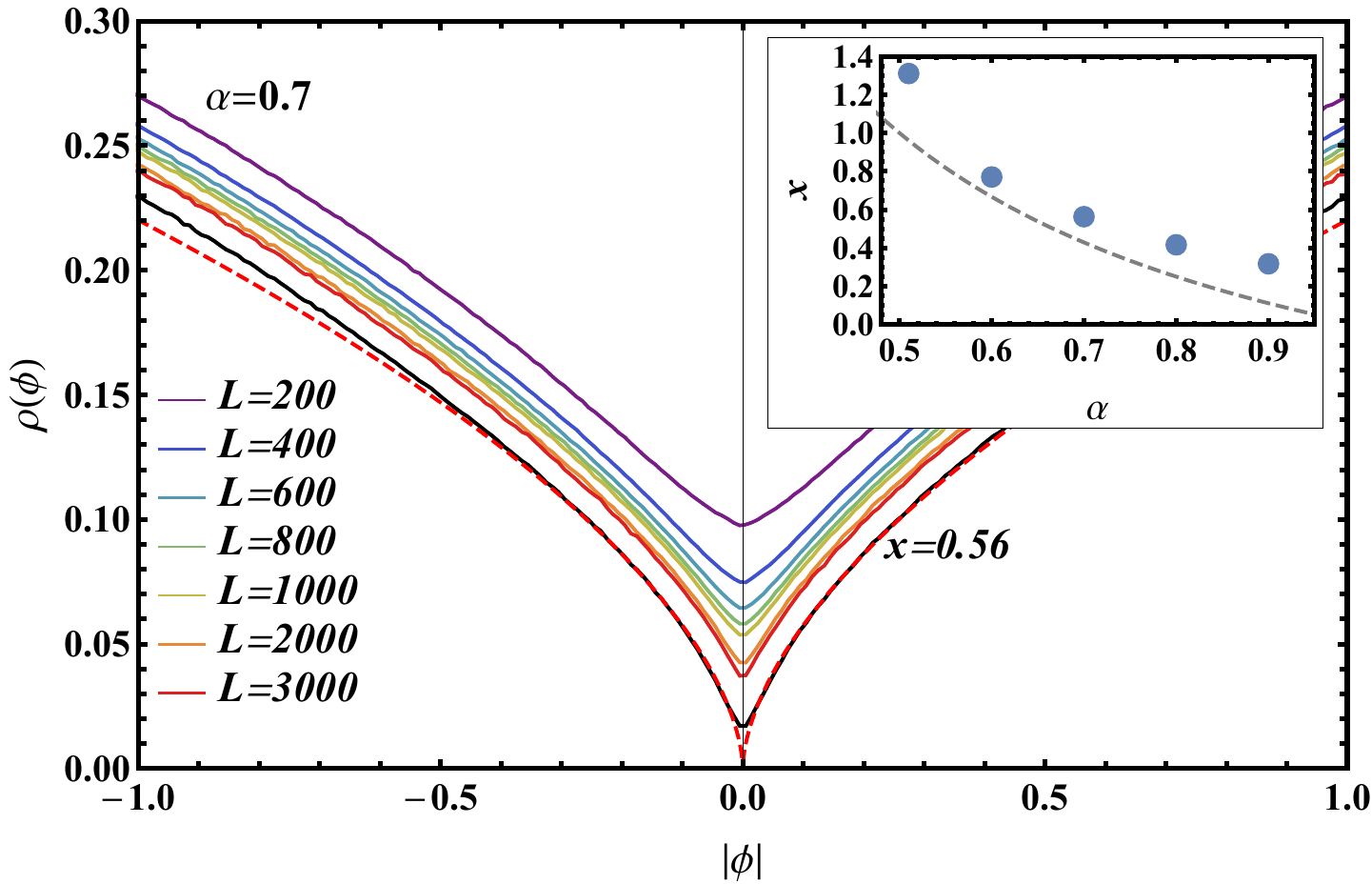}
	\caption{The distribution of effective fields $\phi$ in single-flip stable states, for $\alpha = 0.7$, and various sizes $L$, averaged over 10,000 (for $L=3000$) to 1,000,000 (for $L=200$) disorder realizations. The distributions are extrapolated to $L=\infty$ (black solid line) using $1/L$ scaling up to second order. The $L=\infty$ distribution is fitted for $|\phi| \ll 1$ by a power-law $\rho(\phi) \sim |\phi|^x$. {\bf Inset:} The power $x$ depends on the interaction range parameter $\alpha$. The dashed line represents $\frac{1}{\alpha} - 1$, the threshold for stability against resonance avalanches. }
	\label{Fig:FieldDistribution}
\end{figure}

The distribution of fields for such metastable states for $\alpha = 0.7$ is shown in Fig.~\ref{Fig:FieldDistribution}. In contrast to $T=\infty$ the density of small fields is severely suppressed, following a power-law for sufficiently small fields,
\begin{equation}
	\rho(\phi) \sim |\phi|^x \; \; \; \;  (|\phi| \ll 1).
\end{equation}
This implies that the smallest field found on a system of size $L$ scales as $\phi_{\min} \sim L^{-1/(x+1)}$. We observe that for $L\lesssim 1000$ there are significant finite-size effects of the fields distribution, at least in the range  $|\phi|\lesssim 0.1$. We computed the power $x$ for the single-flip states as a function of $\alpha$, and the result is shown in Fig.~\ref{Fig:FieldDistribution}.

{\bf Resonance avalanches.} Equipped with the distribution of onsite fields, we can now discuss quantum dynamics generated by transverse fields. We will focus on the analysis of resonant processes arising under the application of a (small) transverse field $h_x \ll 1$. Although the analysis has many parallels with the arguments developed by Burin and others~\cite{Burin:2015kz,Gutman16, Maksymov:2017dp,Burin:2015gq,Burin:2017ds, Yao14Dipolar}, the glassiness of the underlying classical problem significantly modifies resonant processes. 

When the system is prepared in an initial product state, only the spins with $|\phi_i|<h_x$ can flip over the time $t\sim 1/h_x$. We will call such spins {\em resonant}. 
The number of resonant spins $N_r(L)$ depends on system size $L$, on the temperature (or energy density) and on the transverse field $h_x$. We denote the typical distance between resonant spins by $d=L/N_r(L)$. 

Turning to dynamics, the resonant spins will oscillate with frequency $\omega\sim h_x$. This, in turn, will affect the effective fields $\phi_j$ felt by other non-resonant spins. The key question is whether these changes can drive new spins to become resonant, and whether such a {\it resonance avalanche} can eventually include the whole system. 

For any (initially) non-resonant spins, the expected change in effective field is $\Delta \phi_j \sim \sum_{i=1}^{N_r(L)} \pm \frac{1}{|d_{i,j}|^\alpha}$
where $d_{i,j}$ is the distance from the $i$th resonant spin to the site $j$. Assuming that there are no correlations in the positions of resonant spins, the typical change of field becomes $(\Delta \phi)^2_{\rm typ} \sim d^{- 2 \alpha}$. If $|\Delta \phi|_{\rm typ} > h_x$, spins that we originally did not count as being resonant can become resonant. It is expected (and verified below) that if this condition is met, flipping one resonant spin will generally cause other spins to become resonant. In this case, the system will exhibit characteristic relaxation dynamics: as resonant spins are flipping, an avalanche of new resonances will lead to a complete loss of spin polarization. We note that such resonance avalanches are reminiscent of the phenomenon of {\it spectral diffusion}\cite{Gutman16,BurinKagan94} in the context of non-glassy models.

The frozen spins can therefore only remain frozen if the resonances do not cause such an avalanche, which requires 
\begin{equation}
	h_x > |\Delta \phi |_{\rm typ} \sim  d^{-\alpha}
	\label{Eqn:ResonantCondition}
\end{equation}
where $d$ is the typical distance between resonant spins.

At infinite temperature $T = \infty$ the distribution of fields is Gaussian, which implies that the average distance between resonant spins is $d \sim h_x^{-1}$. The condition for the stability of spin freezing, Eq.~(\ref{Eqn:ResonantCondition}), becomes $h_x > c h_x^{\alpha}$ with $c$ some $h_x$-independent constant. Because $\alpha <1$, at small $h_x$ this inequality is violated. Therefore, at infinite temperature, we will always have an avalanche of resonances.

However, for low energy density states we found a qualitatively different distribution of fields, $\rho(\phi) \sim |\phi|^x$. Now the distance between resonant spins scales as $d \sim h_x^{-(x+1)}$. Therefore the condition for stability now becomes $h_x > c h_x^{\alpha (x+1)}$, with $c$ an $h_x$-independent constant. This criterion is satisfied for small $h_x$ as long as $\alpha (x+1) > 1$. Therefore, there will be {\em no avalanche of resonances} as long as
\begin{equation}
	x > \frac{1}{\alpha} - 1.
	\label{Eq:xBound}
\end{equation}
This bound is similar to the Efros-Shklovskii bound in the Coulomb glass~\cite{1976JPhC....9.2021E,1975JPhC....8L..49E,Boettcher_2008}. The stability analysis applied to our model yields Eq.~\eqref{Eq:xBound}.

As shown in Fig.~\ref{Fig:FieldDistribution}, this relation is satisfied for the SFS states in our model for any $1/2 < \alpha<1$. While the values of $x$ and $1/\alpha-1$ are close to each other, the bound is not saturated, unlike its analogue in the Coulomb glasses. Such behavior has been observed before in systems, which, similar to our model, had no onsite disorder~\cite{Muller2015,Rademaker2018}. 

\begin{figure}
	\includegraphics[width=0.49\columnwidth]{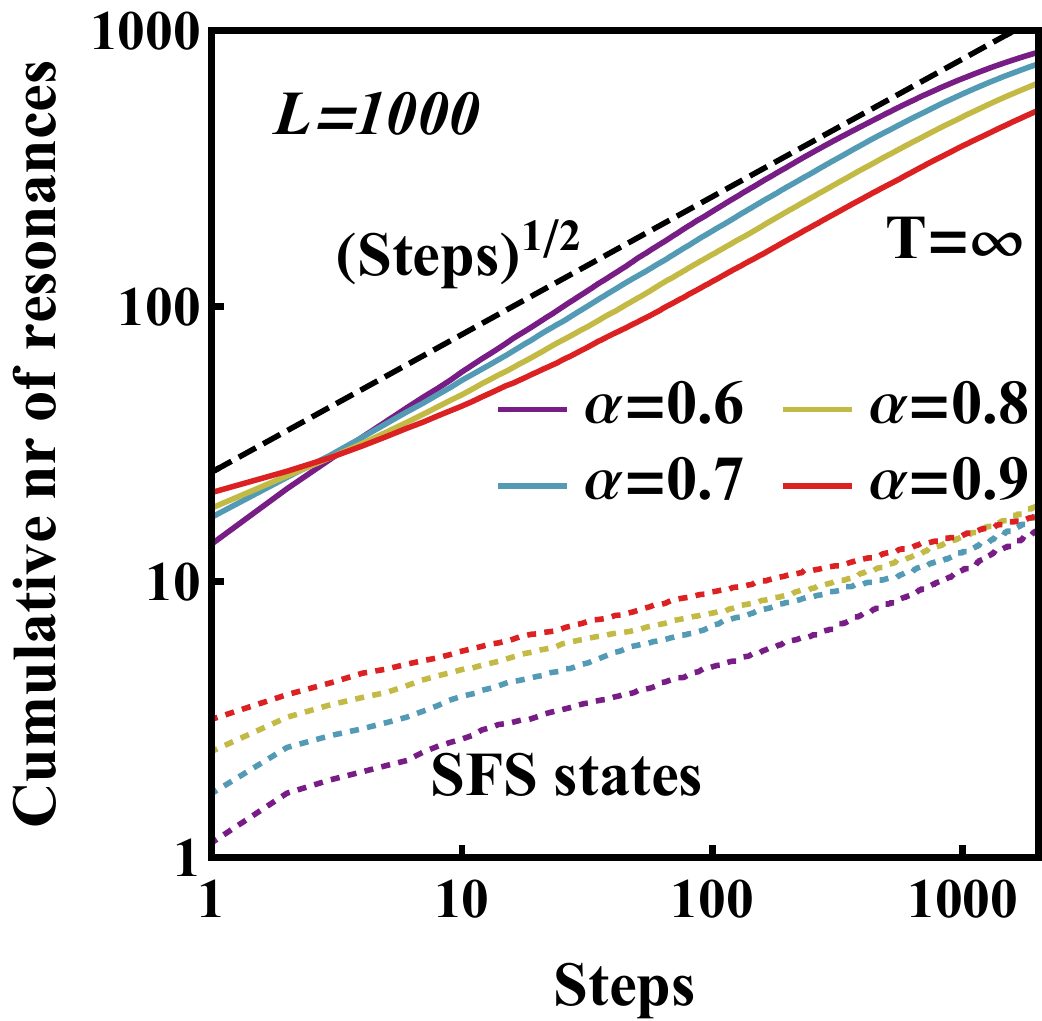}
	\includegraphics[width=0.49\columnwidth]{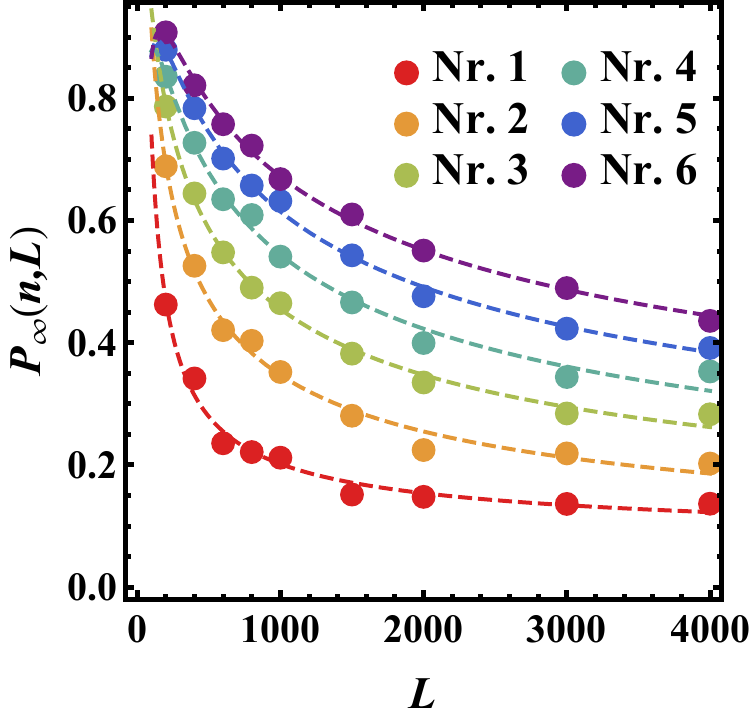}
	\caption{
	\label{Fig:Avalanche}
	{\bf Left:} Number of spins which become resonant during system's evolution. To model the resonance avalanche, at each step (horizontal axis) we randomly flip one resonant spin, keeping track of all the spins that were resonant at some point (vertical axis). Results are shown for system size $L=1000$ averaged over 200 initial states, for varying $\alpha$ and both low energy (SFS) states and $T=\infty$ (random states).
	The high energy density results show an approximately diffusive increase in the total number of resonant spins. 
	For the SFS states, the number of resonant spins for each realization saturates. However, we find that for some realizations saturation occurs at later times than for others, due to the fact that spin flips are performed randomly. This leads to an apparent slow growth of the average number of resonances seen in the Figure. The number of resonant spins, however, was never higher than 2\% of all the spins.
	\label{Fig:ImbalanceFixedSpin}
	{\bf Right:} Remnant spin polarization $P_{\infty}(n,L)$ for the six most resonant spins, as a function of system size $L$\cite{SupplInfo}. The remnant polarizations decay slowly with system size.}

\end{figure}

Next, we test the existence of resonance avalanches in a classical numerical simulation. Given an initial spin configuration, we compute the onsite fields $\phi_i$ and identify all resonant spins that satisfy $|\phi_i | < h_x$. We then flip one of the resonant spins, chosen at random, re-compute the distribution of fields $\phi_i$, and check how many new spins become resonant.  
This is iterated many times. The results, illustrated in Fig.~\ref{Fig:Avalanche}, show that at low energy density the number of resonant spins remains very low within this recursive scheme, suggesting the stability of spin-glass order. At infinite $T$, in contrast, we see as expected a resonance avalanche: the number of spins affected by the avalanche grows approximately as a square root of the number of iterations.

{\bf Two-spin-flip stability.} At the single spin-flip level our analysis shows that for a small field $h_x$ and initial SFS states, most spins remain frozen. However, one can imagine processes involving two spin-flips, where each individual spin flip is not resonant but their combination is. The amplitude of a second-order perturbative correction corresponding to flipping spins $1$ and $2$ equals
\begin{equation}
	A_{12 \rightarrow \bar{1} \bar{2}} = \frac{h_x^2}{- \phi_1 - \phi_2 + 2 \frac{J_{ij}}{r^\alpha_{12}}} \left( \frac{1}{-\phi_1} + \frac{1}{-\phi_2} \right).
\end{equation}
If $|A_{12\to \bar 1\bar 2}|>1$, we will call this process resonant. Now if either $1$ or $2$ are already single spin-flip resonant sites with $\phi_{1,2}<h_x$, the process is naturally accounted for by the resonance avalanches considered above. Our question is thus: how many genuine two spin-flip resonances will exist in this system?

The number of genuine two-spin-flip resonances at $T=\infty$ as a function of system size $L$ can be estimated~\cite{Ho17}, yielding $N_{\rm 2res}(L)\sim L^{2-\alpha}$. We have verified that this approximately holds numerically for systems of size $L\leq 4000$.\cite{SupplInfo} The case of metastable SFS states is more intricate, because of correlations between different spins and the corresponding fields. Our numerical simulations revealed that the number of genuine two-spin-flip resonances grows slower than the system size $L$. This implies that the metastable states are stable with respect to two-flip resonances, consistent with expectations based on the Efros-Shklovskii stability arguments. For initial states that are not only one- but also two-spin flip stable, the number of  two-spin-flip resonances induced by the transverse field is expected to be suppressed even further.


{\bf Quantum dynamics and (non)ergodicity.} Next, we discuss the implications of the resonance avalanches for quench dynamics and eigenstate properties. Since experiments are conducted for finite systems (with tens - hundreds of spins), we will in particular be interested in the effect of finite $L$. 

Let us start with the case of infinite temperature. If $L$ is small such that there are no resonant spins at all (this occurs if $L\lesssim h_x^{-1}$), the system will exhibit usual MBL-like properties, in particular, 
the initial magnetization pattern will fail to relax even at $t\to \infty$, and the system will appear non-ergodic. The eigenstates are also expected to appear MBL-like: in particular, the level statistics is expected to be Poisson. 

Once $L$ is increased such that there are at least a few resonances for a typical initial state, the avalanche will be effective and lead to the decay of initial magnetization. A typical spin will decay after time $t_d\sim 1/h_x^2$, 
but a broad distribution of relaxation times is expected, because spins are gradually included into the avalanche. This provides a direct experimental signature of the resonance avalanche. In this regime, eigenstates at $T=\infty$ are expected to become ergodic, and level statistics will obey Wigner-Dyson distribution.

At low energy density, the non-resonant spins stay non-resonant and will thus retain the memory of their initial magnetization even at very long, and possibly infinite times. Experimentally, this provides a direct signature of ergodicity breaking.

An interesting question concerns the effect of resonant spins on the 'frozen' ones. One possibility is that the resonant spins form a thermal bath which mediates the relaxation of initially non-resonant spins and erasure SG order. As reported below, we have studied the dynamics of the most resonant spins for SFS states using ED, finding indications that this scenario is not realized and SG order remains stable.

We note that SG order may also potentially be destroyed by higher-order, multi-spin resonances, which can e.g. couple different SFS states. In that case, ergodicity may be restored. Given the extreme sparsity of two-spin resonances, we believe this possibility to be unlikely, and this is confirmed by ED studies below.

{\bf Exact diagonalization.} We will now study the model of Eq.~(\ref{MainHamiltonian}) using exact diagonalization (ED) with periodic boundary conditions, which has been used to diagnose MBL phases\cite{Alet:2017dl, AbaninRMP}. Note that the experiments with trapped ions were conducted with similar system sizes $L\approx 10-20$, so results below have direct experimental implications. 

A common tool to distinguish between MBL and ergodic behavior, is to characterize level statistics via the ratio of adjacent eigenvalue gaps, $r = \min(\delta_n, \delta_{n+1}) / \max(\delta_n, \delta_{n+1})$ where $\delta_n = E_n - E_{n-1}$ is the gap between two neighboring eigenvalues. This $r$-value approaches $0.53$ in the ergodic phase, and $r = 0.39$ for Poisson level statistics in the localized phase. 
Additionally, we have studied the Edwards-Anderson order parameter\cite{Edwards:1975ff,stein2013spin} $m_{EA} = \frac{1}{L^2} \sum_{ij} (\langle n | Z_i Z_j | n \rangle)^2$, which tends to zero in the ergodic regime. A finite size scaling up to $L=16$ is included in the Supplementary Information.\cite{SupplInfo}

\begin{figure}
	\includegraphics[width=\columnwidth]{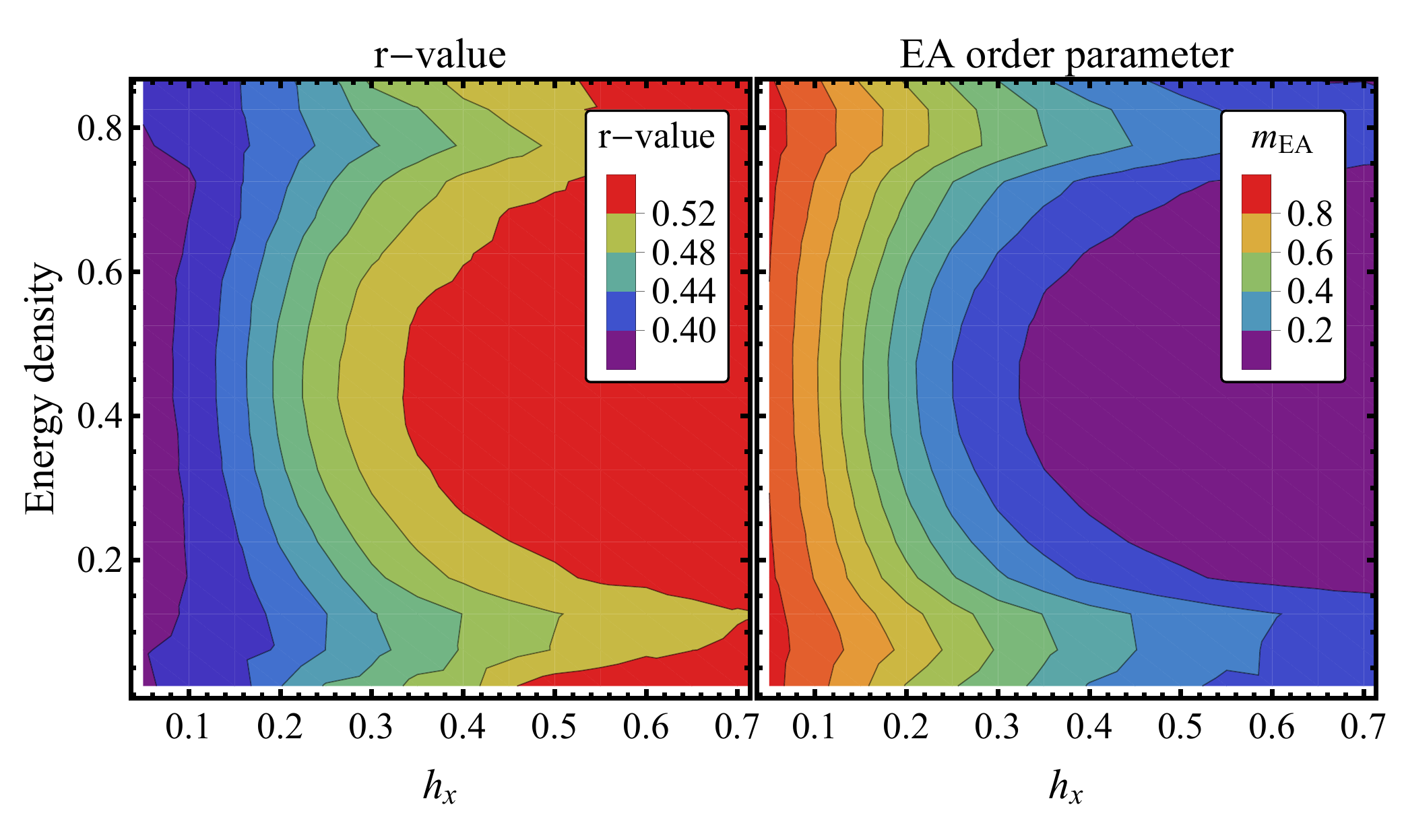}
	\caption{A qualitative phase diagram for $\alpha = 0.7$ can be inferred from two different measures: the $r$-value (left), and the Edwards-Anderson order parameter (right), shown for $L=12$ as a function of scaled energy density ($0$ means ground state and $0.5$ is infinite temperature) and transverse field $h_x$.}
	\label{Fig:RoughPhaseDiagram}
\end{figure}

The two diagnostics, $\langle  r \rangle$ and $m_{EA}$ as a function of energy density $\epsilon = E/L$ and $h_x$ for system size $L=12$ averaged over 2000 disorder realizations, are illustrated in Fig.~\ref{Fig:RoughPhaseDiagram}. The range of interaction is fixed at $\alpha = 0.7$. 

Notably, at very small $h_x\lesssim 0.1$ states at all temperatures appear localized. This is consistent with systems being too small to have resonances. For $h_x \gtrsim 0.2$ where resonance avalanche becomes effective, in the middle of the spectrum ($T = \infty$) the system is clearly ergodic, as the $r$-value approached $0.53$, and the Edward-Anderson order parameter vanishes. At lower energy densities (near the ends of the spectrum) the system stays localized up to larger values of $h_x$.

{\bf Large systems: dynamics of resonant spins.} We further investigate the stability of SG at low energy density by studying the dynamics of most resonant spins in large systems. Here, we make the assumption that spins with fields $|\phi|\gg h_x$ stay frozen. To check the self-consistency of this assumption, we exactly solved the dynamics of the $L_s=6-14$ most resonant spins using ED. Extrapolating this dynamics we estimate the remnant long-time magnetization of the 6 most resonant spins. We find that this magnetization remains sizeable (see Fig.~\ref{Fig:ImbalanceFixedSpin}), slowly decaying to zero as $L\to\infty$, thus establishing that there is no run-away effect of incorporating more and more spins into the exact dynamics of the most resonant spins. Note that while our simulations started with initial SFS states, we expect that our results extend to other low-energy states, because each low energy product state can be made SFS by flipping a small fraction of spins.

Even though the resonant spins do not incite a loss of polarization in the non-resonant spins, amongst themselves they will eventually form an ergodic system. This can be understood by looking at the Hamiltonian for just the resonant spins (labeled by $I,J$): $H_{\rm res}=\sum_I \vec{h}_I \cdot \vec{\sigma}_I + \sum_{I,J} J_{IJ} \frac{{\hat Z}_I {\hat Z}_J}{|I-J|^\alpha}$. The random field $\vec{h}_I=(-h_x, 0, \phi_I)$ has a norm of order $h_x$. The interactions $J_{IJ}$ are now much reduced in strength: given that the typical spacing between resonant spins scales as $d_{\rm res}\sim h_{x}^{-(1+x)}$, the interaction between neighboring resonant spins is of the order $|J_{IJ}| \sim h_x^{\alpha(x+1)}$. As we saw above, $\alpha(x+1)>1$. Thus, the interactions between resonant spins are weaker than the on-site fields, and their long-range nature is expected to lead to eventual thermalization of the resonant-spin subsystem. A complete thermalization, however, requires extremely large system sizes, as incomplete decay of polarization of resonant spins in Fig.~\ref{Fig:ImbalanceFixedSpin} (right) suggests.

{\bf Discussion.} In summary, we proposed to study the interplay of glassiness and MBL -- two generic mechanism of ergodicity breaking -- in a power-law interacting model in 1d. We hope that our work will stimulate experiments with trapped ions, where long-range interactions with a tunable exponent have been demonstrated. 

The onset of glassy behavior leads to an unconventional regime of quantum dynamics: in contrast to high energy density $\epsilon$ where the system behaves as ergodic, at low energy density the memory of initial state is retained. In contrast to MBL systems, a set of resonant spins forms a thermalizing system, which however cannot lead to the decay of SG order. 

We emphasize that the non-ergodic low-$\epsilon$ regime, signalled by the persistence of SG order, is a unique consequence of glassiness: indeed, previous works~\cite{Gutman16} that studied dynamics of long-range interacting systems with power $d<\alpha<2d$, found eventual ergodic behavior (accompanied by diffusive dynamics) even at low $\epsilon$. 

We finally note that it will be interesting to study the high-order tunneling processes between low-energy states separated by large energy barriers -- a problem which has central importance to the performance of the adiabatic quantum algorithm for difficult optimization problems~\cite{AdiabaticGlass}. Such a study will also give an insight into the nature of eigenstates, which may violate the eigenstate thermalization hypothesis, provided the tunneling matrix elements are sufficiently strongly suppressed.
If that is the case, the eigenstates will exhibit clustering similar to that found in infinite-range models~\cite{PalSpinGlass}. 

\acknowledgments \emph{Acknowledgments} - We are thankful to Guido Pagano, Andreas Gei{\ss}ler, Guido Pupillo, Guifre Vidal, Markus M\"{u}ller and Alexander Burin for discussions.
We acknowledge support by Swiss NSF via an Ambizione grant (L.~R.). and a regular project (D.~A.). 

\bibliographystyle{apsrev4-1}

%

\end{document}